\documentclass[aps, twocolumn, showpacs, preprint numbers]{revtex4-1}
\usepackage{amsmath}
\usepackage{graphicx}
\usepackage{hyperref}
\usepackage{epstopdf}
\usepackage{amssymb}
\usepackage{amsfonts}
\usepackage{multirow}

\newcommand{\be}{\begin{equation}} 
\newcommand{\ee}{\end{equation}} 
\newcommand{\bea}{\begin{eqnarray}} 
\newcommand{\eea}{\end{eqnarray}} 
\newcommand{\bqa}{\begin{eqnarray}}
\newcommand{\eqa}{\end{eqnarray}}
\newcommand{\bwt}{\begin{widetext}}
\newcommand{\ewt}{\end{widetext}}
\newcommand{\mb}{\mathbf}

\newcommand{\nn}{\nonumber \\}

\newcommand{\e}{\epsilon}

\newcommand{\w}{\omega}

\begin{document} 
\title {Dynamical Conductivity of Dirac Materials}
\author{Luxmi Rani}\email{luxmiphyiitr@gmail.com, luxmi@prl.res.in}
\author{Navinder Singh}
\affiliation{Theoretical Physics Division, Physical Research Laboratory, Ahmedabad-380009, India.} 
\date{\today}
\begin{abstract} For graphene (a Dirac material) it has been theoretically predicted and experimentally observed that DC resistivity is proportional to $ T^4$ when the temperature is much less than Bloch- Gr\"{u}neisen ($\Theta_{BG}$) temperature and T linear in opposite case ($T>>\Theta_{BG}$). Going beyond the DC case, we investigate the dynamical conductivity in graphene using the powerful method of memory function formalism. In the DC (zero frequency regime) limit, we obtained the above mention behavior which was previously obtained using the Bloch-Boltzmann kinetic equation. In the finite frequency regime, we obtained several  new results: (1) the generalized Drude scattering rate, in the zero temperature limit, shows $\omega^4 $ behavior at low frequencies ($\omega << k_B \Theta_{BG}/ \hbar$) and saturates at higher frequencies. We also observed the Holstein Mechanism, however, with different power laws from that in the case of metals; (2) At higher frequencies, $\omega>>k_B \Theta_{BG}/ \hbar$, and 
higher temperatures $T>>\Theta_{BG}$, we observed that the generalized Drude scattering rate is linear in temperature. In addition, several other results are also obtained. With the experimental advancement of this field, these results should be experimentally tested.
\end{abstract}
\maketitle



\section{Introduction}
  In the recent years, Dirac materials have attracted great attention due to their novel properties which originate from their linear energy dispersion and the presence of Dirac points or lines \cite{JC, TOW}.  
  Graphene is a unique two dimensional Dirac material, consisting of a single atom thick layer of carbon atoms that are closely packed in honeycomb lattice structure due to their $sp^2$ hybridisation \cite{Geim,Novoselov,Zhang,Allen}.
  The electronic band structure of graphene exhibits the high symmetry momentum  points as $\Gamma$, K, $K^{'}$ and M. The points (K and $K^{'}$) where the conduction and valence band touch upon each other are called the Dirac points (showing electronic energy dispersion $E_k=\pm|k|$). The electronic transport in graphene is governed by the relativistic Dirac equation with massless charge carriers (Fermion)\cite{Avouris, Novoselov, Zhang}. The Fermi velocity of charge carriers in graphene is roughly $10^6 m s^{-1}$. Pristine graphene has a minimum ballistic conductivity $\sigma_{
min} =4e^2/h$\cite{Sarma}. 
  
  Among many observable anomalous properties, DC resistivity is the unique one because it is the first to be measured but the last to be understood.  
  The experimentally measured temperature dependent resistivity shows the linear in temperature behavior at high temperatures and $T^{4}$ in the Bloch- Gr\"{u}neisen (BG) regime at low temperature ($T<<\Theta_{BG}$) \cite{Efetov}. The Bloch- Gr\"{u}neisen (BG) temperature is defined in the next section. Recent theoretical investigations\cite{EM, Kaasbjerg, Hwang} have been conducted on graphene by taking into account acoustic phonon limited scattering in the Bloch-Gr\"{u}neisen (BG) regime.
   In ref. \cite{EM} the author analytically solves the Bloch- Boltzmann equation for electron-acoustic phonon scattering, and it is shown that DC resistivity of graphene is proportional to $T^4$ in the low temperature regime ($T<<\Theta_{BG}$) and T-linear in the opposite regime. On similar lines, in ref. \cite{Kaasbjerg, Hwang} the studies of DC resistivity is done within the Boltzmann kinetic theory for the case of electron-acoustic phonon scattering and results agree with the previous investigations mentioned above \cite{Efetov}. 
  
  The problem of DC transport in graphene is also treated using Kubo's linear response theory by several groups \cite{Ziegler, KZ, Gusynin, EGM, Mishchenko, Nomura, Horng, TMR, Radchenko, Falkovsky, Mahan}.  They obtained the universal and nonuniversal observations of minimal conductivity in graphene at frequency $\w\rightarrow0$. In the presence of electric field, Gusynin and Sharapov \cite{Gusynin} analytically obtained the expressions for both the diagonal and off-diagonal conductivities in graphene. It has been observed that these conductivities show the unusual behavior with frequency, chemical potential, and applied field caused by Dirac like spectrum of quasi-particles in this material. 
  
  We now briefly review some recent studies of electronic transport in graphene. Earlier, Radchenko et. al \cite{TMR} have numerically computed the DC conductivity in short range and long range scattering regime for the case of charged impurities \cite{Li, Yan}. Further, the same authors \cite{Radchenko} have studied the effect of charge carriers on the transport properties of graphene within both the Kubo linear response formalism and the Bolch- Boltzmann approach. The  conductivity dependence on carrier density shows sublinear behavior within the semiclassical Boltzmann calculation. However, in contrast to it, the linear behavior is observed within the Kubo formalism.

  Based on the result of Kubo method in the case of the spin-orbital coupling impurity scattering, Liu et. al \cite{Liu} observed that the diagonal conductivity shows a insulating gap at Dirac point, while in opposite case when applying  semiclassical Boltzmann approach, the prediction of insulating gap is failed. It is realized that the low energy transport can not be explained by Boltzmann theory which occurs from the spin orbit coupling scattering. 
  
  Several experimental groups have studied the conductivity measurement in different frequency regimes (visible, infrared, near infrared and ultraviolet) \cite{Novoselov, WZ, NMR, LI, T. Stauber, Bolotin, Heo, Park}. From experimental investigations, it has been observed that conductivity versus carrier density shows linear to sublinear behavior. Experimentally, Zhang et. al \cite{WZ} obtained the AC conductivity $\sigma (\w)$ in tetrahertz (THz) frequency range 0.2-1.2 THz. At present,  $\sigma (\w) \propto \w^4$ is not observed.
  
  However, theoretical investigations of dynamical conductivity due to electron-acoustic phonon scattering in graphene is lacking. In this paper, we present our results of the calculation of dynamical conductivity of graphene using the memory function formalism. The memory function formalism was pioneered by Mori and Zwanzig \cite{Mori,Zwanzig}. This is one of the most useful method known as, projection operator method. The basic idea of this method is to study the time correlation function systematically in the case of many body correlated systems. G\"otze and  W\"olfle \cite{GW} put this method on a more useful footing by introducing their perturbative scheme and 
  used that to solve the electrical conductivity in simple metals by taking into account different interactions. We use the G\"otze - W\"olfle method to investigate the dynamical conductivity of graphene.

 This paper is organized as follows. In section \ref{sec:memory}, we will discuss the memory function formalism to calculate the dynamical
conductivity in graphene. In section \ref{sec:result}, the results and discussion of dynamical conductivity due to electron-acoustic phonon scattering in graphene will be explained analytically and numerically. Finally, we will summarize our results and present our conclusions.

\section{Memory function Formalism}
\label{sec:memory}
Our aim is to study the dynamical conductivity of a two dimensional (2D) Dirac material such as graphene using the Memory function formalism. Following the previous work \cite{GW, Singh, Kubo}, we are going to use an alternative representation of conductivity from Kubo's formalism i.e.
\begin{eqnarray} 
\sigma(z, T)&=&i\frac{\chi_0}{z - M(z, T)}.
\end{eqnarray} 
Here M(z) is the memory function. $z=\w+i\delta$ is the complex frequency and we have taken the limit $\delta\rightarrow0$. $ \chi_0$ represents the static limit of correlation function (i.e. $ \chi_0 = Ne/m $). First, we will calculate the memory function. The memory function can be expressed through the current-current correlation function:
\begin{eqnarray}  
 \chi(z)&=&\langle\langle j_1;j_1\rangle\rangle = i \int_{0}^\infty e^{izt}\langle [j_1, j_1]\rangle dt.
 \end{eqnarray}
Here inner $\left<......\right>$ describes canonical ensemble averages and outer $\left<......\right>$ represents Laplace transform of the expectation value of the same. It was shown in \cite{GW, Singh, Pankaj, Das} that the correlation function can be written as:
 \be
 \chi(z)=\frac{\langle\langle C;C\rangle\rangle_{z=0}-\langle\langle C;C\rangle\rangle_z}{z^2}.
 \ee
 Where $C=[j_1, H]$. H is the total Hamiltonian of the system. And the memory function can be written as \cite{GW, Singh, Pankaj, Das};
 \be
 M(z,T)\simeq \frac{z\chi(z)}{\chi_0}.
 \ee
  Using the above relation, we obtain an approximate expression for the memory function
 \bea
 M(z, T)=\frac{\langle\langle C;C\rangle\rangle_{z=0}-\langle\langle C;C\rangle\rangle_z}{z\chi_0}.
 \eea
 Defining $\phi(z)=\langle\langle C;C\rangle\rangle_{z}$,  the memory function can be written as,
 \bea 
 M(z, T)= \frac{\left[\phi(0)-\phi(z)\right]}{(z\chi_0)}.
 \label{eq:M}
 \eea
In the present work, we consider a two dimensional dirac material having linear isotropic energy dispersion $(\epsilon_k=\hbar v_f|\vec{k}|)$, where $k$ is the momentum and $v_f$ is the Fermi velocity. Also in the whole analytical calculation we consider the system of units where $\hbar=1$ and $k_B=1$.
 Thus, to study the memory function for Dirac material with electron-phonon scattering \cite{Ziman}, we write the following Hamiltonian,
 \bea
 H&=&H_{electrons} + H_{electron-phonon}
 \eea
 Here
\bea
H_{e^-}&= &\sum_{\mb{k}\sigma} \epsilon_k c^\dagger_{\mb{k}\sigma}c_{\mb{k}\sigma}\\
H_{e^--ph}&=&\sum_{\mb{k},\, \mb{k'},\sigma}\left[ D(\mb{k}-\mb{k'})c^\dagger_{\mb{k}\sigma}c_{\mb{k'}\sigma}b_{\mb{k}-\mb{k'}} + H.\, c.\right].
 \label{eq:H} 
\eea
Here, $c^\dagger_{\mb{k}\sigma} (c_{\mb{k}\sigma})$ is the creation (annihilation) operator having spin $\sigma$ and momentum k. $b^\dagger_{\mb{k}-\mb{k'}}(b_{\mb{k}-\mb{k'}})$ is the phonon creation (annihilation) operator having phonon wave vector $q=\mb{k}-\mb{k'}$.
The coefficient $D(\mb{k}-\mb{k'})$ in the electron phonon Hamiltonian is given by the expression for acoustic phonon's \cite{EM,Sarma}i.e. 
\bea
D(q)&=&-i \bigg(\frac{\hbar}{2\rho_m \w_q} \bigg)^ {1/2} \times D_0q \left[1-\bigg(\frac{q}{2k_f}\bigg)^2\right]^{1/2}
\label{eq:e-ph-mat}
\eea
Here, $D_0$ is the deformation potential coupling constant for graphene, $ \rho_m $ is surface mass density. The phonons are considered as longitudinal accoustic having a dispersion of the form $ \w_q = v_s q $. Here $ v_s $ is the sound velocity, and q is the phonon momentum. In the present case, we have the condition \cite{EM} $ \frac{\hbar \w_q}{k_B}\leq min (\Theta _{BG}, \Theta_D) $ with $\Theta_D$ is the Debye temperature, and $\Theta _{BG} =2(\frac{v_s}{v_f})\frac{\epsilon_f}{k_B}$ the Bloch Gr\"{u}neisen temperature. $\e_f$ is the Fermi energy. For our computation, we need to compute $C$ and the correlator $\langle\langle C;C\rangle\rangle_z$. For this, we use the expression for electrical current as $j=\Sigma_{k\sigma}v_x(k)c^\dagger_{\mb{k}\sigma}c_{\mb{k}\sigma}$, where $v_x (k)$ is the velocity of carriers with momentum $\mb{k}$ in the x- direction. Now from the electron-phonon interaction part of the Hamiltonian (Eqn.\ref{eq:H}) we get, 
\be
 C= \sum_{k,k'}[v_x(\mb{k})-v_x(\mb{k'})][D(\mb{k}-\mb{k'})c^\dagger_{\mb{k}\sigma} c_{\mb{k'}\sigma}b_{\mb{k}-\mb{k'}}-H.c.].
 \ee 
With this expression, two particle correlator $\phi(z)=\langle\langle C; C\rangle\rangle$ can be computed as follows,
 \bea
\phi(z)=-\sum_{k,k',\sigma}\sum_{p,p'}[v_x(\mb{k})-v_x(\mb{k'})][v_x(\mb{p})-v_x(\mb{p'})]\nn D(\mb{k}-\mb{k'}) D^\star(\mb{p}-\mb{p'}) \nn 
\left\langle\left\langle c^\dagger_{\mb{k}\sigma} c_{\mb{k'}\sigma}b_{\mb{k}-\mb{k'}};b^\dagger_{\mb{p}-\mb{p'}}c^\dagger_{\mb{p'}}c_{\mb{p}}\right\rangle\right\rangle\nn
 -\sum_{k,k',\sigma}\sum_{p,p'}[v_x(\mb{k})-v_x(\mb{k'})][v_x(\mb{p})-v_x(\mb{p'})] \nn D^\star(\mb{k}-\mb{k'})D(\mb{p}-\mb{p'})\nn  \left \langle\left\langle c^\dagger_{\mb{k'}\sigma} c_{\mb{k}\sigma}b^\dagger_{\mb{k}-\mb{k'}};b_{\mb{p}-\mb{p'}}c^\dagger_{\mb{p}}c_{\mb{p'}}\right\rangle\right\rangle .
\eea
Further, simplifications of the correlation function $\langle\langle c^\dagger_{\mb{k}\sigma} c_{\mb{k'}\sigma}b_{\mb{k}-\mb{k'}};b^\dagger_{\mb{p}-\mb{p'}}c^\dagger_{\mb{p'}}c_{\mb{p}}\rangle\rangle$ leads to \cite{Singh}:
\bea
  -\frac{1}{\epsilon_{k}-\epsilon_{k'}-\omega_{k-k'}+z}[f_k (1-f_{k'})(1+n_{\omega_{k-k'}})\nn
-f_k'(1-f_k)n_{\omega_{k-k'}}]\delta_{p,k}\delta_{p,k'}.
\eea
Here $f_k =\langle c^\dagger_{\mb{k}}c_{\mb{k}}\rangle$,  $f_k'=\langle c^\dagger_{\mb{k'}}c_{\mb{k'}}\rangle$ are the Fermi-Dirac distribution functions and, $n_{\omega_{k-k'}}=\langle b^\dagger_{\mb{k}-\mb{k'}}b_{\mb{k}-\mb{k'}}\rangle$ is the Bose-Einstein distribution function. Similarly the second factor of $\phi(z)$ can be simplified with the combine result
\bea
\phi(z)=\sum_{k,k'}\left|D(\mb{k}-\mb{k'})\right|^2[v_x(\mb{k})-v_x(\mb{k'})]^2 \nn
 \times[{f (1-f')(1+n_{-}))-f'(1-f)n_{-}}]\nn
\times \left[\frac{1}{(\epsilon_{k}-\epsilon_{k'}-\omega_{-}+z)}+ \frac{1}{(\epsilon_{k}-\epsilon_{k'}-\omega_{-}-z)}\right].
\eea
Here we have used the notation $f=f_k$; $f'= f_k'$; $n_{-}=n_{\omega_{k-k'}}$; and $\omega_{-}=\omega_{k-k'}$.

Using the above expression in Eqn.\ref{eq:M}, the imaginary part of the memory function can be written as,
\bea
ImM(\w, T)& = &\frac{\pi}{\chi_0}\sum_{kk'} \left|D(\mb{k}-\mb{k'})\right|^2[v_x(\mb{k})-v_x(\mb{k'})]^2\nn
&&\times[(1-f)f'n_{-}]\nn
&&\times\Bigg[\frac{(e^{\beta\omega}-1)}{\w}\delta(\epsilon_k-\epsilon_{k'}- \omega_{-}+\w) -\nn
&&..(\rm terms\, with\, \w\rightarrow -\w)\Bigg].
\label{eq:ImM}
\eea
After simplification, the final memory function expression becomes
\bea
ImM(\w, T)&=&M_0\int_0^{q_{BG}} dq \times\frac{q^3}{k_f}\sqrt{1-({q}/{2k_f})^2}\nn
&&\times \frac{1}{e^{\beta\w_q}-1}
\times\Bigg\lbrace(1-\frac{\w_q}{\w})\frac{e^{\beta\w}-1}{(e^{\beta(\omega-\w_q)}-1)}\nn
&&+(\rm terms\, with\, \w\,\rightarrow -\w)..\Bigg\rbrace.
\label{eq:ImMf}
\eea

The details of the calculations are given in Appendix [\ref{sec:relation}]. Here, $q_{BG}$ being the  Bloch- Gr\"{u}neisen momentum. As discussed in Appendix [\ref{sec:relation}], in the graphene case, the the radius of Fermi circle is smaller as compare to Debye circle. In the scattering process, we need to consider those phonons whose wave vector is less than the Fermi wave vector. Therefore, the upper cut off values  of q integral is to be $2k_f$. Further, to obtain the information from the above expression, we analysed it in special limiting cases that are discussed in the next section \ref{sec:result}. Consider first the zero frequency limit.

\section{Results And Discussion}
\label{sec:result}
 \textbf{Case-I}: DC limit case (at zero frequency)\\
 Consider $\w\rightarrow0$ then in the (Eqn.\ref{eq:ImMf}), curly bracket term will become
 \bea
 &&\left\lbrace(1-\frac{\w_q}{\w})\frac{e^{\beta\w}-1}{(e^{\beta(\omega-\w_q)}-1)}+(\rm terms\, with\, \w\,\rightarrow -\w)..\right\rbrace \nn 
 &&\approx \frac{2\beta\w_q}{1-e^{-\beta\w_q}}
 \eea
 With this expression, the imaginary part of the memory function (Eqn.\ref{eq:ImMf}) can be written as,
\bea
ImM(0)&=&2M_0 \int_0^{q_{BG}} dq \times \frac{q^3}{k_f}\sqrt{1-({q}/{2k_f})^2}\nn &&\times\frac{1}{e^{\beta\w_q}-1}\times\frac{\beta\w_q}{1-e^{-\beta\w_q}}
 \eea
 Using  $x=\beta\w_q$ and $ \beta v_s q_{BG}= \beta k_B{\Theta_{BG}}=\frac{\Theta_{BG}}{T}$ and $q_{BG}\approx 2k_f$, we obtain
\bea
ImM(0)&=&\frac{2M_0 }{k_f\beta^4 v_s^4}\int_0^{\beta v_s q_{BG}} dx \times \frac{{x^4}\sqrt{1-({x}/{2k_f\beta v_s})^2}}{(e^{x}-1)(1-e^{-x})}\nn
&=&\frac{2M_0 }{k_f}\left(\frac{T}{\Theta_{BG}}\right)^4{q^4_{BG}}\nn &&\times\int_0^{\frac{\Theta_{BG}}{T}} dx \frac{{x^4}\sqrt{1-({x}/{(\Theta_{BG}/T)})^2}}{(e^{x}-1)(1-e^{-x})}  \nn
&=&\frac{2M_0 }{k_f}\left(\frac{T}{\Theta_{BG}}\right)^4{q^4_{BG}} \int_0^{z}  \frac{{x^4}\sqrt{1-({x}/{z})^2}}{(e^{x}-1)(1-e^{-x})}dx \nn
&=&\frac{2M_0 }{k_f}\left(\frac{T}{\Theta_{BG}}\right)^4{q^4_{BG}}\times{\mathcal{J}}_{4}(z)
 \eea
 Here, we have the following function\cite{EM}:
\bea
{\mathcal{J}}_{n}(z) =
\int_{0}^{z}\frac{x^{n}\sqrt{1-(x/z)^{2}}}{(1-e^{-x})(e^{x}-1)}dx
\eea
The asymptotic limit of this function, for $n>0$ an integer, is given by
${\mathcal{J}}_{n}(\infty) = n!\zeta(n)$,
with $\zeta(n)$ the Riemann Zeta function. In particular, for $n=4$, it has the values $\zeta(4) = \pi^{4}/90$ .\\
Subcase(1): for low temperature behavior $T<<\Theta_{BG}$,  $z\rightarrow\infty$, 
${\mathcal{J}}_{n}(\infty) = 4!\zeta(4)$. Thus,
\bea
ImM(0)&=&\frac{2M_0 }{k_f}\left(\frac{T}{\Theta_{BG}}\right)^4{q^4_{BG}}\times\frac{4!\pi^{4}}{90}\nn
 ImM(0)&\propto &T^4.
\eea
Subcase(2): for high temperature behavior $T>>\Theta_{BG}$, $z\rightarrow zero$. Then ${\mathcal{J}}_{4}(z) \approx \frac{\pi}{16} \times z^3$ (see Appendix \ref{sec:rel}).

Therefore, we find that the imaginary part of memory function at high temperature becomes
\bea
ImM(0)&=&\frac{2M_0 }{k_f}\left(\frac{T}{\Theta_{BG}}\right){q^4_{BG}}\times\frac{\pi}{16} \nn
ImM(0)&\propto &T.
\eea
These results are in agreement with the existing theoretical and experimental   data reported in literature \cite{Efetov, EM, Hwang, Kaasbjerg}.

\textbf{Case-II:} Finite frequency regime (at zero temperature)

In $T\rightarrow0$ limit case, the ultralow behavior can be analyzed in Eqn.\ref{eq:ImMf} at all frequencies. Thus the expression in curly bracket leads to $ \left(1-\frac{\w_q}{\w}\right)e^{\beta \w_q}$ at $\w>\w_q$, and $ \left(\frac{\w_q}{\w}-1 \right)e^{\beta \w}$ at $\w<\w_q$.
\bea
ImM(\w)&=&M_0\int_0^{q_{BG}}dq\times\frac{q^3}{k_f}\sqrt{1-({q}/{2k_f})^2}\nn
&&\times\frac{1}{e^{\beta\w_q}-1}\Theta(\w-\w_q)\left(1-\frac{\w_q}{\w}\right)e^{\beta \w_q} \nn 
&&+ M_0\int_0^{q_{BG}}dq\times\frac{q^3}{k_f}\sqrt{1-({q}/{2k_f})^2}\nn
&&\times\frac{1}{e^{\beta\w_q}-1}\Theta(\w_q-\w)\left(\frac{\w_q}{\w}-1 \right)e^{\beta \w}.\\
ImM(\w)&=&M_0\int_0^{\w}d\w_q\times\frac{{\w_q}^3}{k_f v_s^4}\sqrt{1-({\w_q}/{2k_f v_s})^2}\nn
&&\times\frac{e^{\beta\w_q} }{e^{\beta\w_q}-1}\bigg(1-\frac{\w_q}{\w}\bigg)\nn 
&+& M_0\int_\w^{\w_{BG}}d\w_q\times\frac{{\w_q}^3}{k_f v_s^4}\sqrt{1-({\w_q}/{2k_f v_s})^2}\nn
&&\times\frac{e^{\beta\w}}{e^{\beta\w_q}-1}\bigg(\frac{\w_q}{\w}-1 \bigg).
\eea
Here we consider $\w_{BG}=2v_s k_f$ and $x' =\w_q$,  then rewrite the above equation as:
\bea
ImM(\w)&=& \frac{2M_0}{v_s^3} \int_0^{\w}dx'\times\frac{{x'}^3}{\w_{BG}}\sqrt{1-({x'}/{\w_{BG}})^2}\nn
&&\times\frac{e^{\beta x'} }{e^{\beta x'}-1}\bigg(1-\frac{x'}{\w}\bigg)\nn
&+& \frac{2M_0}{v_s^3}\int_\w^{\w_{BG}}dx' \times\frac{{x'}^3}{\w_{BG}}\sqrt{1-({x'}/{\w_{BG}})^2}\nn &&\times\frac{e^{\beta\w}}{e^{\beta x'}-1}\bigg(\frac{x'}{\w}-1 \bigg).
\label{eq:ImM1}
\eea
At $T\rightarrow 0$, the first term of Eqn.\ref{eq:ImM1} provides the dominating contribution (because second term is exponentially small). Thus, the imaginary part of the memory function becomes\\
\bea
ImM(\w)&=& \frac{2M_0}{v_s^3} \int_0^{\w}dx'\times\frac{{x'}^3}{\w_{BG}}\sqrt{1-({x'}/{\w_{BG} })^2}\bigg(1-\frac{x'}{\w}\bigg)\nn 
\eea
Let $x'/\w_{BG}= y$, thus we have
\bea
 ImM(\w)&=& \frac{2M_0 \w_{BG}^3}{v_s^3} \Bigg[\int_0^{\frac{\w}{\w_{BG}}}dy \times y^3\sqrt{1-y^2} -\nn
 &&\frac{\w_{BG}}{\w} \int_0^{\frac{\w}{\w_{BG}}}dy \times y^4\sqrt{1-y^2})\Bigg]
\eea
At $\w<<\w_{BG}$, On solving the integral, we obtain
 \bea
 ImM(\w)&=& \frac{M_0 \w^4}{10 v_s^3 \w_{BG}}\nn
 ImM(\w) &\propto & \w^4.
 \eea
  Thus $\w^4$ dependence of scattering rate is obtained in given Dirac material case. However, in metals, it shows the $\w^5$ dependence.

 In the case $\w>>\w_{BG}$, $T\rightarrow 0$, the imaginary part of the memory function is 
 \bea
 ImM(\w)&=& \frac{2M_0}{v_s^3} \int_0^{\w_{BG}}dx'\times\frac{{x'}^3}{\w_{BG}}\sqrt{1-({x'}/{\w_{BG} })^2}\nn
 &&\times\frac{e^{\beta x'} }{e^{\beta x'}-1}\bigg(1-\frac{x'}{\w}\bigg)
 \eea
Again let $x'/\w_{BG}= y$, thus we have
\bea
 ImM(\w)&=& \frac{2M_0 \w_{BG}^3}{v_s^3} \Bigg[\int_0^{1}dy \times y^3\sqrt{1-y^2} -\nn
 &&\frac{\w_{BG}}{\w} \int_0^{1}dy \times y^4\sqrt{1-y^2})\Bigg]
 \eea
On solving the integral in the above limits, we obtain,
\bea
ImM(\w)&=& \frac{M_0 \w_{BG}^3 }{ v_s^3 }\Bigg[\frac{4}{15}-\left(\frac{\w_{BG}}{\w}\right)\frac{\pi}{16}\Bigg]\nn
ImM(\w)&=& \Bigg[a-\frac{b}{\w} \Bigg].
\eea
Here, $a=\frac{4}{15}\frac{M_0 \w_{BG}^3 }{ v_s^3 }$ and $ b=\frac{\pi M_0 \w_{BG}^4 }{16 v_s^3 } $. Thus in the high frequency limit it saturates.

 \textbf{Case-III:} Finite frequency regime (at finite temperature)
 
 Consider $\w>>\Theta_{BG}, T$ then in Eqn.\ref{eq:ImMf}, curly bracket term can be simplified as
 \bea
 &&\left\lbrace(1-\frac{\w_q}{\w})\frac{e^{\beta\w}-1}{(e^{\beta(\omega-\w_q)}-1)}+(\rm terms\, with\, \w\,\rightarrow -\w)..\right\rbrace \nn 
 &&\approx 2(e^{\beta\w_q}+1)
  \eea
In this case the Eqn.\ref{eq:ImMf} becomes
 \bea
 ImM(\w, T)&=&\frac{2M_0}{k_f}\int_0^{q_{BG}}q^3\sqrt{1-\left(\frac{q}{2k_f}\right)^2} \frac{(e^{\beta\w_q}+1)}{(e^{\beta\w_q}-1)}dq\nn
&=&\frac{2M_0}{k_f}\int_0^{q_{BG}}q^3\sqrt{1-\left(\frac{q}{2k_f}\right)^2}\coth \bigg({\frac{\beta \w_q}{2}}\bigg)dq\nn
 \eea
 Using  $x=\beta\w_q$ and $ \beta v_s q_{BG}= \beta k_B{\Theta_{BG}}=\frac{\Theta_{BG}}{T}$.
 \bea
ImM(\w, T)&=&\frac{2M_0}{k_f}\left(\frac{T}{\Theta_{BG}}\right)^4{q_{BG}^4} \times\int_0^{\frac{\Theta_{BG}}{T}} dx \times {x^3}\nn
&& \times\sqrt{1-({x}/{(\Theta_{BG}/T)})^2}
\times \coth\left({\frac{x}{2}}\right)\nn
&=&\frac{2M_0}{k_f}\left(\frac{T}{\Theta_{BG}}\right)^4{q_{BG}^4} \times J_3\left(\frac{\Theta_{BG}}{T}\right).
\label{eq:ImMf1}
\eea
Here $J_3(z)=\int_0^{z} dx \times {x^3}\sqrt{1-({x}/{z})^2}\times \coth\left({\frac{x}{2}}\right)$.

In the low temperature case: $T<<\Theta_{BG}$, $z\rightarrow\infty$, $J_3(z)=\frac{z^4}{4}$ the resulting  Eqn.\ref{eq:ImMf1} is
\bea
ImM(\w, T)&=&\frac{M_0}{2k_f}{q_{BG}^4}\nn
ImM(\w, T)&\propto & constant.
\eea
In the high temperature case: $T>>\Theta_{BG}$, $z\rightarrow\ 0$, $J_3(z)=\frac{\pi z^3}{8}$ the resulting  Eqn.\ref{eq:ImMf1} is
\bea
ImM(\w, T)&=&\frac{\pi M_0}{4k_f}{q_{BG}^4} \left(\frac{T}{\Theta_{BG}}\right)\nn
ImM(\w,T) &\propto& T.
\eea
Hence the scattering rate is linearly dependent on temperature in high temperature case  when $T>>\Theta_{BG}$ and $\w>>\Theta_{BG}, T$. All the above mentioned results in different limiting cases are given in the  table \ref{T:results}.

\begin{center}
\begin{table}
\begin{tabular}{|l|l|l|}
\hline
	$\w$ and $T$ regime & $ImM(\w)$\\
\hline
 \hline
	$\w= 0;\,\,\,\, T<<\Theta_{BG}$ & $T^4.$\\
\hline
	$\w= 0;\,\,\,\, T>>\Theta_{BG}$  & $ T$. \\
\hline
	$\w<<\w_{BG}; \, \, \, \, T\rightarrow0$ & $ \w^4$.\\
\hline
	$\w>>\w_{BG}; \, \, \, \,  T\rightarrow0$ & $ [a- \frac{b}{\w}] $.\\
\hline
	$\w>> \Theta_{BG}, T; \, \, \,\, T<<\Theta_{BG}$ & $ \propto constant. $\\
\hline
	$\w>> \Theta_{BG}, T; \, \, \, \,T>>\Theta_{BG}$ & $ \propto T.$\\
\hline
\end{tabular}
\caption{The results of imaginary part of the memory function in different limiting cases.}
 \label{T:results}
 \end{table}
\end{center}

 In the general case, we need to perform the integral in  Eqn.\ref{eq:ImMf} numerically. With $\hbar =6.582 \times 10^{-16} eV.s $ and Boltzmann constant $ k_B = 8.617 \times10^{-5}eV/K$, we perform the integral by setting frequency in energy units (in eV). Now we present our numerical results.

\begin{figure}[htbp!]
        \centering
        \includegraphics[angle=0,width=7.5cm]{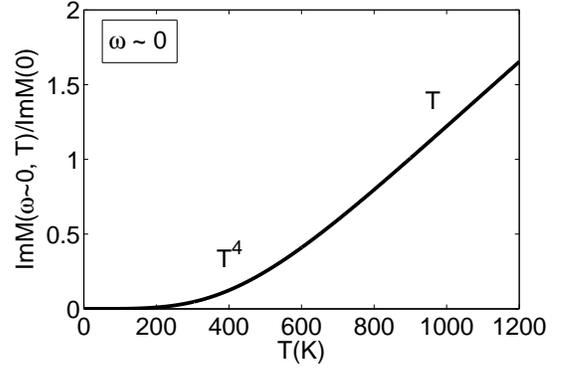}
           \caption{Variation of the imaginary part of the memory function with temperature at zero frequency.}
           \label{fig:figimmt}
	\end{figure}
\begin{figure}[htbp!]
        \centering
        \includegraphics[angle=0,width=7.5cm]{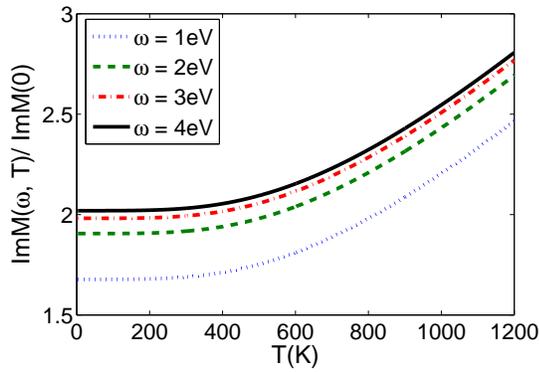}
           \caption{Variation of the imaginary part of the memory function with temperature at different frequencies.}
           \label{fig:figimm-tt}
	\end{figure}

In Fig.\ref{fig:figimmt}, we show the variation of the imaginary part of the scaled memory function with temperature at zero frequency (ImM(0)=$\frac{ N^2 D_0^2}{32\pi^3\chi_0\rho_m v_s}$). From  Fig.\ref{fig:figimmt}, we observe that ImM(0, T) shows $T^4$ behavior at very low temperatures ($T << \Theta_{BG}$) and T-linear behavior at high temperatures ($T >>\Theta_{BG}$). These features are in agreement with experimental data \cite {Efetov}. In Fig.\ref{fig:figimm-tt}, we depict the variation of $ImM(\w, T)$ as a function of temperature at different frequencies. It shows that on increasing the frequency, scattering rate ($ImM(\w)$) shifts upward. It is also observed that scattering rate shows linear behavior at high temperature, while at low temperature it shows saturation. We notice that even at zero temperature there is finite scattering. This is nothing but the Holstein mechanism for the present case.

\begin{figure}[htbp!]
        \centering
         \includegraphics[angle=0,width=7.5cm]{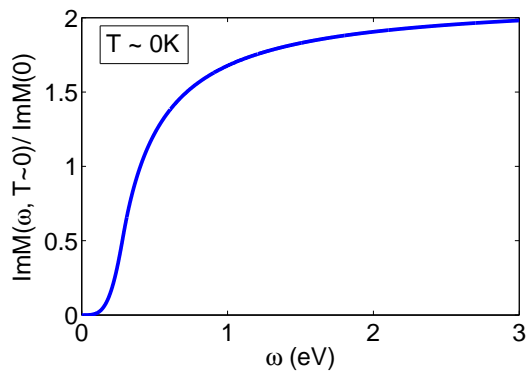}
           \caption{Variation of the imaginary part of the memory function with frequency at zero temperature for graphene.}
           \label{fig:figimm-mw}
	\end{figure}

\begin{figure}[htbp!]
        \centering
        \includegraphics[angle=0,width=7.5cm]{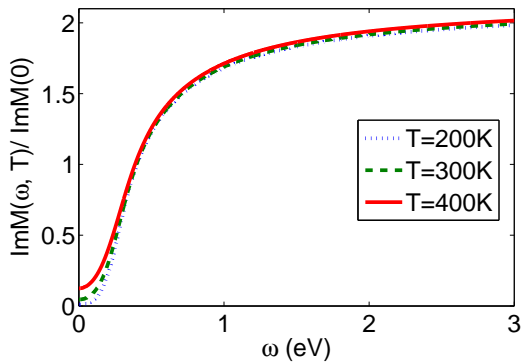}
           \caption{Variation of the imaginary part of the memory function with frequency at different temperatures.}
           \label{fig:figimm}
	\end{figure}

 The frequency dependent behavior of imaginary part of memory function at zero temperature is shown in Figure \ref{fig:figimm-mw}. From Fig.\ref{fig:figimm-mw}, we observed that in the low frequency limit ($\hbar\w<<k_B\Theta_{BG}$), $ImM(\w, 0)/ImM(0)$ shows $\w^4$  behavior. While, at high frequency limit it becomes constant. In Fig.\ref{fig:figimm}, we depict the variation of imaginary part of memory function with frequency at various temperatures. It is observed that the $ImM(\w, T)/ImM(0)$ increases with the increase in temperature at lower frequency regime while at the higher frequencies, it shows saturation behavior.
   
\section{Conclusions}
\label{sec:conclusion}
 We have presented a theoretical study of dynamical conductivity of a Dirac material by using the memory function formalism with electron- acoustic phonon scattering. We have reproduced the zero frequency (DC) results namely, $T^4$ behavior of resistivity at low temperature and T-linear at high temperature \cite{Efetov,EM, Sarma, Kaasbjerg}. 
 
 The Holstein mechanism \cite{TH} which is the finite absorption of electromagnetic energy at zero temperature is also predicted by our calculations in 2D Dirac materials. However with different scaling behavior (Table \ref{T:results}). In the Holstein mechanism a photon generates an electron-hole pair along with the emission of a phonon (i.e. phonon assisted scattering). We predict that the scattering increases as $\w^4$ in low temperature limit (refer to Table \ref{T:results}).

 Both at high frequencies and at high temperatures $(T>> \Theta_{BG})$, we have observed the linear behavior of memory function with temperature, while at low temperature scattering rate $(ImM(\w, T))$ becomes constant. These  temperature and frequency dependent results should be experimentally tested.

\section*{Acknowledgement}
We thank to Nabyendu Das and Pankaj Bhalla for useful suggestions and discussions.

\appendix
\section{Derivation of Eqn.\ref{eq:ImMf}}
\label{sec:relation}
In the present study, we consider graphene (as a Dirac material), having linear isotropic energy dispersion $(\epsilon_f=v_f|\vec{k}|)$. To simplify equation (\ref{eq:ImM}), we insert $[v_x(\mb{k})-v_x(\mb{k'})]^2=v_f^2(1-cos{\theta})$ and convert the sums into integrals in Eqn.\ref{eq:ImM}. We further reduced it by considering that $k$ is pointing along the $x$-direction and $k'$ subtends an angle $\theta$ with it. Also we insert an integrated delta function$(1=\int dq\delta(q-|\mb{k}-\mb{k'}|))$ over $q$ and resulting expression is
\bea
ImM(\w)&=&\frac{\pi N^2 v_f^2}{(2\pi)^4\chi_0}\int_0^\infty dq D_q^2\int_0^\infty {kdk} \int_0^\infty {k'dk'}\nn
&&\int_0^{2\pi} d\theta (1-cos{\theta})\delta(q-\left|\mb{k}-\mb{k'}\right|)\nn
&&\times[(1-f)f'n_{-}]\nn &&\times\Bigg[\frac{(e^{\beta\omega}-1)}{\w}\delta(\epsilon_k-\epsilon_{k'}-\omega_{-}+\w) -\nn 
&&..(\rm terms\, with\, \w\,\rightarrow -\w)\Bigg].
\eea
Shift the integration variables to $\e_k=v_fk$, $\e_k'=v_fk'$, $d\e=v_fdk$. And for our convenience, we use the notation $\e_k=\e$ and $\e_k'=\e'$.
\bea
\Rightarrow\int_0^\infty {kdk} \int_0^\infty {k'dk'}=\frac{1}{v_f^4}\int_0^\infty {\e d\e} \int_0^\infty {\e'd\e'}\nn
\delta(q-\left|\mb{k}-\mb{k'}\right|)=\delta(q-\sqrt{k^2+k'^2-2kk'\cos\theta})
\eea
\bea
ImM(\w)&=&\frac{\pi N^2}{(2\pi)^4\chi_0 v_f^2}\int_0^\infty dq D_q^2\int_0^\infty {\e d\e} \int_0^\infty {\e' d\e'}\nn 
&&\int_0^{2\pi} d\theta (1-cos{\theta})
\delta(q-\frac{1}{v_f}\sqrt{\e^2+\e'^2-2\e\e'\cos\theta})\nn
&& \times[(1-f)f'n_{-}]\nn
&&\times\bigg[\frac{(e^{\beta\omega}-1)}{\w}\delta(\e-\e'-\omega_{-}+\w)-...\nn
&&(\rm terms\, with\,\w\,\rightarrow -\w)...\bigg]  
\eea
Where, $\theta$ is the scattering angle. We are considering a degenerate system,  and the magnitude of Fermi energy is much greater than $k_B T$. Thus, in graphene, electron-acoustic phonon scattering occurs in a very thin layer close to the Fermi surface. Within this $\e\approx\e'\approx \e_f$, the integral over $\theta$ can be simplified as follows:
\bea
I_\theta&=&\int_0^{2\pi} d\theta (1-cos{\theta})\delta(q-\frac{1}{v_f}\sqrt{\e^2+\e'^2-2\e\e'\cos\theta})\nn
&=&\int_0^{2\pi} d\theta (1-cos{\theta})\delta(q-k_f\sqrt{2(1-\cos\theta)})\nn
&=&\int_0^{\pi} d\theta (1-cos{\theta})\delta(q-k_f\sqrt{2(1-\cos\theta)})\nn &+&\int_\pi^{2\pi} d\theta (1-cos{\theta})\delta(q-k_f\sqrt{2(1-\cos\theta)})
\eea
 Here change to $x=1-cos{\theta}$
\bea
I_\theta&=&\int_0^{2} dx \frac{x}{\sqrt{2x-x^2}}\delta(q-k_f\sqrt{2x})\nn &-&\int_2^{0} dx \frac{x}{\sqrt{2x-x^2}}\delta(q-k_f\sqrt{2x})\nn
&=&2\int_0^{2} dx \frac{x}{\sqrt{2x-x^2}}\delta(q-k_f\sqrt{2x})\nn
&=&2\int_0^{2k_f}dt\frac{t^3}{2k_f^4}\frac{1}{\sqrt{\frac{t^2}{k_f^2}(1-\frac{t^2}{4k_f^2})}}\delta(q-t)\nn
&=&2\frac{q^2}{2k_f^3}\frac{1}{\sqrt{(1-\frac{q^2}{4k_f^2})}}
=\frac{q^2}{k_f^3}\frac{1}{\sqrt{(1-\frac{q^2}{4k_f^2})}}
\eea
Substituting this value the $\theta$-integral into the memory function approximation. Thus the imaginary part of the memory function can be written as
\bea
ImM(\w)&=&\frac{\pi N^2}{(2\pi)^4\chi_0}\int_0^\infty {d\e} \int_0^\infty {d\e'}\int_0^\infty dq \times D_q^2\times\frac{q^2}{k_f}\nn
&&\times\frac{1}{\sqrt{(1-\frac{q^2}{4k_f^2})}}\times \frac{1}{(e^{\beta\e}+1)}\frac{1}{(e^{\beta\e'}+1)}\nn
&&\times\frac{1}{(e^{\beta\w_q}-1)}
\Bigg[\frac{(e^{\beta\omega}-1)}{\w} \delta(\epsilon-\epsilon'-\omega_{-} +\w) 
-\nn
&&(\rm terms\, with\, \w\,\rightarrow -\w)..\Bigg]
\eea
Now performing the $\e'$ integral using the property of the delta functions. The resulting memory function expression becomes,
\bea
ImM(\w)&=&\frac{\pi N^2}{(2\pi)^4\chi_0}\int_0^\infty {d\e}\int_0^\infty dq D_q^2\times\frac{q^2}{k_f}\frac{1}{\sqrt{(1-\frac{q^2}{4k_f^2})}}\nn
&&\times\frac{1}{e^{\beta\e}+1}\frac{1}{e^{\beta\w_q}-1}\nn
&&\times\Bigg[\frac{(e^{\beta\omega}-1)}{\w}\times\frac{1}{e^{\beta\e}{e^{\beta(\w-\w_q)}}+1}\nn
&&-..(\rm terms\, with\, \w\,\rightarrow -\w)..\Bigg].
\eea
And, then solving the integration over $\e$ using the elementary method $\int_{-\infty}^{+\infty}dx \frac{e^x}{e^{x}+1} \frac{1}{e^{x+a}+1} $. The integral becomes
\bea
I_\e&=&\int_0^\infty {d\e}\frac{e^{\beta\e}}{e^{\beta\e}+1} \frac{1}{e^{\beta\e+\beta(\w-\w_q)}+1}\nn
&=&\frac{(\w-\w_q)}{e^{\beta(\w-\w_q)}-1}
\eea
And thus the imaginary part of memory function is
\bea
ImM(\w)&=&\frac{\pi N^2}{(2\pi)^4\chi_0}\int_0^{q_{BG}} dq\times D_q^2\times\frac{q^2}{k_f}\frac{1}{\sqrt{(1-\frac{q^2}{4k_f^2})}}\nn
&&\times\frac{1}{e^{\beta\w_q}-1} \times\Bigg\lbrace\frac{(e^{\beta\omega}-1)}{\w}\frac{(\w-\w_q)}{e^{\beta(\w-\w_q)}-1}\nn
&&+(\rm terms\, with\, \w\, \rightarrow -\w)\Bigg\rbrace
\eea
Using the relation $D_q$ from Eqn.\ref{eq:e-ph-mat},  defining $M_0=\frac{ N^2 D_0^2}{32\pi^3\chi_0\rho_m v_s}$, and $\w_q=v_sq$, we have
\bea
ImM(\w)&=&M_0\int_0^{q_{BG}} dq \times\frac{q^3}{k_f}\sqrt{1-({q}/{2k_f})^2}\frac{1}{e^{\beta\w_q}-1}\nn
&&\times\Bigg\lbrace(1-\frac{\w_q}{\w})\frac{e^{\beta\w}-1}{(e^{\beta(\omega-\w_q)}-1)}+\nn
&&(\rm terms\, with\, \w\,\rightarrow -\w)..\Bigg\rbrace.
\eea
\section{Some useful relations}
\label{sec:rel}
We have used the following relation:
\bea
{\mathcal{J}}_{n}&=&\int_{0}^{z}\frac{x^{4}\sqrt{1-(x/z)^{2}}}{(1-e^{-x})(e^{x}-1)}dx
\eea
For small z case, let $x=z\cos(\theta)$ then above integral becomes
\bea
{\mathcal{J}}_{n}&=&\int_{0}^{\pi/2}\frac{z^{5}\cos^4(\theta)\sin^2(\theta)e^{z\cos(\theta)}}{(e^{z\cos(\theta)}-1)^2}d\theta\nn
&\approx & \frac{\pi z^3}{16}
\eea

\end{document}